\documentclass[review,12pt]{elsarticle}

\usepackage{amsmath,amssymb}
\usepackage{graphicx}
\usepackage{xcolor}
\usepackage{booktabs}
\usepackage{dcolumn}
\usepackage[colorlinks=true,
            linkcolor=blue,
            citecolor=blue,
            urlcolor=blue]{hyperref}

\journal{Mechanics of Materials}

\begin{document}

\begin{frontmatter}

\title{Microscopically informed constitutive framework for stress overshoot, yielding, and strain hardening in amorphous materials}

\author[unimi]{Ankit Singh}
\ead{ankit.singh@unimi.it}

\author[msu]{Valeriy V. Ginzburg}
\ead{ginzbur7@msu.edu}

\author[unimi]{Alessio Zaccone\corref{cor1}}
\ead{alessio.zaccone@unimi.it}

\affiliation[unimi]{organization={Department of Physics ``A. Pontremoli'', University of Milan},
            addressline={Via Celoria 16},
            city={Milan},
            postcode={20133},
            country={Italy}}

\affiliation[msu]{organization={Department of Chemical Engineering and Materials Science, Michigan State University},
            city={East Lansing},
            state={Michigan},
            postcode={48824},
            country={USA}}

\begin{abstract}
We develop a microscopically informed constitutive framework for the nonlinear deformation of metallic and polymer glasses based on nonaffine elasticity coupled to irreversible many-body relaxation. The framework predicts the full stress-strain response, from linear elasticity through stress overshoot and yielding to steady plastic flow. We show that stress overshoot originates from the competition between a nonaffine elastic instability induced by strain-driven loss of mechanical connectivity at the atomic/molecular level, and viscous dissipation associated with structural relaxation. For polymer glasses, finite chain extensibility naturally accounts for strain hardening at large deformation. The stretched-exponential relaxation exponent is obtained independently from stress or modulus relaxation measurements and provides the primary dynamical input to the constitutive framework. Using a small set of physically meaningful parameters, the model quantitatively reproduces experimental stress-strain curves for metallic glasses, polycarbonate, PMMA, and epoxy resins over a broad range of strain rates. These results establish a unified constitutive framework linking relaxation dynamics, yielding, plastic flow, and strain hardening in amorphous solids.
\end{abstract}


\end{frontmatter}

\section{Introduction}
Understanding how amorphous solids deform under load remains a central challenge because elasticity, yielding, and plastic flow emerge in the absence of the crystalline defects that govern deformation in crystals \cite{ferry1961viscoelastic,siviour2016high,choi2026solid,hertzberg2020deformation}. At temperatures above their brittle-ductile transition (BDT), the glassy polymers \cite{siviour2016high} such as polycarbonate (PC), poly(methyl methacrylate) (PMMA) \cite{arruda_1995_PMMAeffects,dupaix_2007constitutive}, epoxy resin \cite{jordan2008mechanical,colak_2019_epoxy}, and highly entangled coarse grained (CG) polymer melt \cite{nelson2026shear,hoy_2007strain_hardenig,JCP_2025} exhibit a characteristic nonlinear stress-strain behavior, where an initially elastic response is followed by a pronounced yield point at small strains of a few percent. Beyond yielding, the stress typically drops (strain softening) or saturates depending on thermal and mechanical history \cite{meijer2003multi}, signaling the onset of plastic flow mediated by irreversible microscopic rearrangements. At sufficiently large strains, many systems display strain hardening \cite{siviour2016high, hem2022microscopic,zhao_2024_hardenig_modeling_microscopic}, where the stress increases again with deformation, a feature crucial for mechanical stability and resistance to failure \cite{wang2012elastic,falk1998dynamics,hoy2008strain}.

Strain hardening is a defining feature of many polymer glasses and plays a crucial role in suppressing strain localization and delaying catastrophic failure \cite{meijer2005mechanical,van2003origin}. It is commonly associated with chain entanglements, cross-linking, and molecular orientation under deformation, giving rise to hardening moduli of order $10^6$--$10^8$ Pa below the glass transition temperature \cite{haward2012physics}. The strain-hardening modulus is several orders of magnitude lower than the initial elastic modulus (which, for the glassy polymers, is on the order of 1-10 GPa). While highly entangled polymers typically exhibit pronounced post-yield hardening and enhanced ductility, weakly entangled systems often fail in a brittle manner \cite{siviour2016high,hem2022microscopic}. Despite extensive study, the microscopic origin of strain hardening and its connection to yielding and plastic flow remain subjects of ongoing debate \cite{siviour2016high,hem2022microscopic}.

A similar phenomenology is observed in metallic glasses, which exhibit yielding, stress overshoot, and steady-state flow despite the absence of crystalline order \cite{lu2003deformation,PhysRevLett.108.098303}. Unlike crystalline solids, where plasticity is governed by dislocation motion as established in the seminal works of Orowan, Polanyi, and Taylor \cite{orowan_1934,polanyi1934lattice,taylor1934mechanism}, amorphous solids deform through spatially localized irreversible rearrangements. These events are commonly described as shear transformation zones (STZs), which act as elementary carriers of plastic flow \cite{argon1979plastic,falk1998dynamics,falk2011deformation,henits_2012free}. STZs generate long-ranged elastic fields that interact through Eshelby-like stress redistribution and can organize into correlated deformation patterns \cite{eshelby1957determination,picard2004elastic}. STZ theory further describes
plasticity as the collective activation and evolution of localized rearranging
zones, with their transformation kinetics controlled by the local structural
state, effective temperature, and free volume
\cite{falk_2011deformation,kamrin_2014two,voyiadjis_2016constitutive}. Despite their success in describing plastic flow, the microscopic connection between localized rearrangements, elastic instability, and macroscopic stress overshoot remains incompletely understood \cite{nicolas2018deformation,jiang_2015origin}. More recently, mean-field STZ theories have been developed specifically for
amorphous polymers, incorporating the evolution of STZ density and orientation,
effective temperature, and polymer-chain stretching to describe yielding,
strain hardening, loading-path dependence, and the Bauschinger
effect~\cite{lin2023mean,lin2026three}. The recent three-dimensional extension
further provides a finite-deformation formulation in which chain stretching
modifies the STZ transformation kinetics.  STZ theory describes localized rearrangements and plastic-flow kinetics.  In contrast, the present nonaffine framework addresses the structural origin of the elastic instability through strain-induced loss of mechanical connectivity and the associated nonaffine softening and the
stress overshoot.

A microscopic perspective is provided by nonaffine particle
displacements, which play a central role in the elastic response of disordered
solids~\cite{dong_2023non,wang_2022PRL,Zaccone_2011,laurati_2017long}. In this framework, local force imbalances generated by deformation drive nonaffine relaxations that reduce the elastic free energy, while irreversible cage rearrangements govern plastic flow \cite{koumakis2012yielding,koumakis2016start}. Here we show that the interplay between these two processes provides a microscopic description of nonlinear deformation in glasses. Specifically, we demonstrate that stress overshoot arises from a nonaffine elastic instability caused by strain-induced loss of mechanical connectivity competing with viscous relaxation. The resulting constitutive relation quantitatively describes metallic and polymer glasses and naturally captures yielding, plastic flow, and strain hardening within a single framework.

\section{Constitutive framework}
We consider the free energy of deformation of an amorphous solid under shear strain $\gamma$, written as $F = F_A(\gamma) - F_{NA}(\gamma)$, where $F_A$ and $F_{NA}$ denote the affine and nonaffine contributions, respectively, and $\gamma$ is the applied shear strain. The affine contribution $F_A$ corresponds to the standard elastic energy, as described by the Born-Huang theory \cite{born1996dynamical}, in which all particles deform uniformly following the applied macroscopic strain. In contrast, disordered systems lack long-range symmetry, leading to additional nonaffine displacements that reduce the overall free energy \cite{Zaccone_2011,PhysRevLett.110.178002}.

The nonaffine contribution $F_{NA}$ originates from the absence of local inversion symmetry in disordered solids \cite{zaccone2023}. Under deformation, particles experience unbalanced forces that cannot be compensated by affine motion alone. The resulting nonaffine displacements relax these forces and lower the elastic free energy, leading to a reduction of the Young modulus relative to the Born prediction \cite{Zaccone_2011,zaccone2023}.

In densely packed systems such as metallic glasses or polymer glasses, the effective interactions can be described using a potential of mean force $V_{\mathrm{eff}}$, related to the radial distribution function via $V_{\mathrm{eff}}/k_BT = -\ln g(r)$ \cite{hansen1988theory}. The coordination number $n_b$ is then determined by the number of neighbors within the first coordination shell near the distance $R_0$. Under strain, the local cage surrounding each particle becomes distorted, leading to a strain-dependent coordination number $n_b(\gamma)$. In extension regions, particles move apart and bonds are lost, while in compression regions, the formation of new contacts is limited by excluded-volume effects, resulting in a net decrease of $n_b$ with increasing strain. This progressive loss of nearest-neighbor connectivity weakens the local cage and reduces the number of mechanically active constraints that can sustain the applied load. As the coordination decreases, the elastic stiffness is progressively reduced, leading to nonaffine softening of the material and ultimately to the elastic instability associated with the maximum in the stress-strain curve.

The evolution of the coordination number can be described by a shear strain- and rate-dependent form using the relation for the strain-dependent mean number of mechanically-active bonds $n_b$ derived in \cite{PhysRevB_2014_MG}: $n_b(\gamma) = \frac{n_b^0}{2}\left(1 + e^{-A\gamma}\right)$, where $A = \Delta/k_BT + 1/(\dot{\gamma}\tau_c)$ incorporates both thermal activation and strain-induced cage breaking. Here, $\tau_c$ is the structural relaxation time at fixed strain rate $\dot{\gamma}$ and $\Delta \sim k_B T_g$ represents the energy barrier for cage rearrangement. At large strain, $n_b \to n_b^0/2 \approx 6$ with $n_b^0 \approx 12$ as for dense amorphous matter \cite{hansen1988theory} such as metallic glass, reflecting the loss of mechanical stability at the isostatic point $n_b =6$ (for materials where bonding is mostly central-force and non-covalent) and the transition to fluid-like behavior \cite{Zaccone_2011}. Thus, the elastic instability can be interpreted microscopically as the point at which the strain-induced loss of local connectivity brings the system close to isostatic point. For polymers, assuming an 8-chain lattice Arruda-Boyce model for the unstrained polymer \cite{arruda1993evolution}, one would go from $n_b^0 = 8$ to the isostatic point for polymers (estimated accounting for both covalent and van der Waals-type interactions) $n_b^0/2 = n_b^c=4$ as theoretically predicted in \cite{PhysRevLett.110.178002}.
Substituting $n_b(\gamma)$ into the free energy of deformation $F_{\mathrm{el}}=\frac{1}{2}K\left[n_b(\gamma)-n_b^{c}\right]\gamma^{2}$, where \(n_b^{c}\) is the critical coordination number corresponding to the marginal stability or isostatic point, and differentiating with respect to shear strain $\gamma$ yields a nonlinear stress-strain relation,
\begin{equation}\label{stress_el}
\sigma_{\mathrm{el}}(\gamma)=
\frac{1}{4}
n_b^0 K \gamma
e^{-\gamma\left(\frac{T_g}{T}+\frac{1}{\dot{\gamma}\tau_c}\right)}
\left[
2-\gamma
\left(
\frac{T_g}{T}
+
\frac{1}{\dot{\gamma}\tau_c}
\right)
\right],
\end{equation}
where $K = \frac{2}{5\pi}(\kappa \phi / R_0)$ is an effective elastic constant.

Here, we reformulate the uniaxial specialization within finite-deformation kinematics. Let $\lambda_{\varepsilon}$ denote the axial stretch and let $\varepsilon=\ln\lambda_{\varepsilon}$ be the corresponding signed axial Hencky (logarithmic) strain. For proportional uniaxial loading we write the deformation gradient as
\begin{equation}
\mathbf{F}=
\begin{pmatrix}
\lambda_{\varepsilon} & 0 & 0\\
0 & \lambda_{\perp} & 0\\
0 & 0 & \lambda_{\perp}
\end{pmatrix},
\qquad
J=\det\mathbf{F}=\lambda_{\varepsilon}\lambda_{\perp}^{2},
\end{equation}
where $\lambda_{\perp}$ is the transverse stretch. For the proportional loading paths considered here, the measured Poisson ratio is used as an effective logarithmic transverse-contraction parameter, $\ln\lambda_{\perp}=-\nu\varepsilon$. Since $\mathbf{F}$ is diagonal, the right stretch tensor is $\mathbf{U}=(\mathbf{F}^{T}\mathbf{F})^{1/2}$ and the Hencky strain tensor $\mathbf{H}=\ln\mathbf{U}$ becomes
\begin{equation}
\mathbf{H}=
\begin{pmatrix}
\varepsilon & 0 & 0\\
0 & -\nu\varepsilon & 0\\
0 & 0 & -\nu\varepsilon
\end{pmatrix}.
\end{equation}
The distortional part is $\mathbf{H}'=\mathbf{H}-\frac{1}{3}\mathrm{tr}(\mathbf{H})\mathbf{I}$. We first define the non-negative equivalent distortional magnitude $\Gamma_{\mathrm{eq}}=\sqrt{2\,\mathbf{H}':\mathbf{H}'}$
and the corresponding signed distortional Hencky measure
\begin{equation}
\Gamma=
\Gamma_{\mathrm{eq}}\,\mathrm{sign}(\varepsilon)
=
\sqrt{2\,\mathbf{H}':\mathbf{H}'}\,\mathrm{sign}(\varepsilon)
=
\frac{2(1+\nu)}{\sqrt{3}}\,\varepsilon .
\label{signed_Gamma}
\end{equation}
Thus $\Gamma$ changes sign continuously with the axial Hencky strain, whereas $\Gamma_{\mathrm{eq}}$ represents only the amount of distortional deformation. For $\nu=0.5$, Eq.~\eqref{signed_Gamma} reduces to $\Gamma=\sqrt{3}\,\varepsilon$. The factor $2(1+\nu)/\sqrt{3}$ follows directly from the tensor norm $\sqrt{2\,\mathbf{H}':\mathbf{H}'}$.

To use a single stress measure for all constitutive contributions, we employ the equivalent Cauchy (von Mises) stress magnitude
\begin{equation}
q=\sqrt{\frac{3}{2}\,\mathbf{s}:\mathbf{s}},
\qquad
\mathbf{s}=\boldsymbol{\sigma}-\frac{1}{3}\mathrm{tr}(\boldsymbol{\sigma})\mathbf{I},
\end{equation}
where $\boldsymbol{\sigma}$ is the Cauchy stress tensor, and we associate with it the signed scalar stress $\sigma=q\,\mathrm{sign}(\varepsilon)$.

For a uniaxial stress state with traction-free lateral surfaces this signed scalar reduces to the axial Cauchy stress $\sigma_{11}$. Likewise, for simple shear the corresponding signed equivalent stress is $\sqrt{3}\,\tau$. The constitutive equations below are therefore written in signed form; the conventional positive magnitudes used for the compression plots are introduced only at the end of the constitutive development.

The finite-deformation rate measure is obtained from the velocity gradient $\mathbf{L}=\dot{\mathbf{F}}\mathbf{F}^{-1}$ and the rate-of-deformation tensor $\mathbf{D}=\frac{1}{2}(\mathbf{L}+\mathbf{L}^{T})$. Its non-negative equivalent distortional magnitude is
\begin{equation}
\dot{\Gamma}_{\mathrm{eq}}=\sqrt{2\,\mathbf{D}':\mathbf{D}'},
\qquad
\mathbf{D}'=\mathbf{D}-\frac{1}{3}\mathrm{tr}(\mathbf{D})\mathbf{I}.
\end{equation}
For a monotonic proportional loading branch, let $s=\mathrm{sign}(\dot{\varepsilon})$, which is constant and equals $\mathrm{sign}(\varepsilon)$ after loading starts. Then
\begin{equation}
\dot{\Gamma}_{\mathrm{eq}}
=
\frac{2(1+\nu)}{\sqrt{3}}\,s\dot{\varepsilon}>0 .
\end{equation}
The cage-breaking kinetics depend on the deformation magnitude $\Gamma_{\mathrm{eq}}$ and on the positive equivalent rate $\dot{\Gamma}_{\mathrm{eq}}$, whereas the sign of the stress is carried by the signed measure $\Gamma$. This preserves the original monotonic-loading physics without inserting absolute-value signs into the signed kinematics.

With these definitions, the elastic contribution to the signed equivalent Cauchy stress is
\begin{equation}
\label{elastic_uniaxial}
\sigma_{\mathrm{el}}(\varepsilon)
=
\frac{\sqrt{3}}{4}\,
n_b^0 K\,
\Gamma\,
\exp\!\left[
-\Gamma_{\mathrm{eq}}
\left(
\frac{T_g}{T}
+
\frac{1}{\dot{\Gamma}_{\mathrm{eq}}\tau_c}
\right)
\right]
\left[
2-
\Gamma_{\mathrm{eq}}
\left(
\frac{T_g}{T}
+
\frac{1}{\dot{\Gamma}_{\mathrm{eq}}\tau_c}
\right)
\right].
\end{equation}
Because $\Gamma$ is signed while the cage-breaking factor depends on the equivalent magnitude $\Gamma_{\mathrm{eq}}$, Eq.~\eqref{elastic_uniaxial} gives the same stress magnitude on the monotonic tensile and compressive branches with the appropriate opposite sign. In the limit $\varepsilon\rightarrow0$, the initial response is unique and linear, $\sigma_{\mathrm{el}}\simeq E\varepsilon$, with $E=n_b^0K(1+\nu)$. Thus the original microscopic nonaffine instability is retained while being embedded in a finite-deformation Hencky-strain/Cauchy-stress setting.

To simplify the notation, we define the dimensionless parameter
\begin{equation}
A=
\frac{T_g}{T}
+
\frac{1}{\dot{\Gamma}_{\mathrm{eq}}\tau_c}
=
\frac{T_g}{T}
+
\frac{\sqrt{3}}
{2(1+\nu)\,s\dot{\varepsilon}\tau_c},
\label{A_def}
\end{equation}
where $s=\mathrm{sign}(\dot{\varepsilon})$ is fixed on a monotonic loading branch, so that $s\dot{\varepsilon}>0$. The parameter $A$ therefore combines the thermal activation and finite-deformation equivalent Hencky strain-rate contributions governing the strain-induced loss of mechanical connectivity.
The strain-rate dependence entering through the parameter
$A$ reflects the kinetics of deformation-induced loss of mechanical
connectivity rather than an additional equilibrium thermodynamic state
variable. Accordingly, the elastic stress represents the instantaneous
nonequilibrium response of the evolving microscopic network, whereas the
irreversible dissipation associated with structural relaxation is described
separately through the viscous contribution. We therefore do not interpret
the complete constitutive relation as deriving from an equilibrium Helmholtz
free-energy potential, nor do we claim a general hyperelastic formulation
valid for arbitrary loading histories. The present framework is a nonequilibrium
constitutive framework specialized to monotonic constant-rate deformation,
with the rate dependence entering through the structural relaxation kinetics. A
fully general thermodynamic internal-variable formulation, in which the coordination number evolves as an independent state variable subject to an appropriate dissipation inequality, is beyond the scope of the present work. Nevertheless, this interpretation is consistent with the broader physical picture emerging from microscopic nonequilibrium theories based on coupled structural and stress evolution \cite{mutneja_2026microscopic,mutneja_2026unified}.

The elastic contribution to the signed equivalent Cauchy stress can then be written compactly as
\begin{equation}
\sigma_{\mathrm{el}}(\varepsilon)=
\frac{\sqrt{3}}{4}
n_b^0 K\,
\Gamma\,
e^{-A\Gamma_{\mathrm{eq}}}
\left[
2-
A\Gamma_{\mathrm{eq}}
\right].
\label{elastic_compact}
\end{equation}
Equation~\eqref{elastic_compact} naturally predicts an elastic instability corresponding to the maximum stress magnitude on either monotonic loading branch. The characteristic distortional yield scale is $\Gamma_{\mathrm{eq},y}\sim A^{-1}$; using $\Gamma_{\mathrm{eq}}=2(1+\nu)s\varepsilon/\sqrt{3}$, this corresponds to an axial Hencky-strain scale of order $\sqrt{3}/[2(1+\nu)A]$.

In addition to the elastic response, dissipation arises from microscopic friction and structural relaxation processes. For proportional uniaxial loading, the axial component of $\mathbf{D}$ is the Hencky strain rate, $D_{11}=\dot{\varepsilon}=\dot{\lambda}_{\varepsilon}/\lambda_{\varepsilon}$. The viscous contribution is therefore written as a signed Cauchy-stress contribution driven by the signed Hencky strain and strain rate. For monotonic constant-rate deformation, $\varepsilon$ and $\dot{\varepsilon}$ have the same sign and hence $\varepsilon/\dot{\varepsilon}=t\geq0$. The generalized Maxwell/KWW form can therefore be written without absolute-value signs as \cite{Bingyu,laurati_2017long,klages2008anomalous},
\begin{equation}
\sigma_{\mathrm{visc}}(\varepsilon)
=
\dot{\varepsilon}\eta
\left[
1-\exp\!\left\{
-\left(
\frac{\varepsilon}{\dot{\varepsilon}\tau_v}
\right)^{\beta}
\right\}
\right],
\end{equation}
where $\eta$ is the viscosity, $\tau_v$ is the viscoelastic relaxation time, and $\beta$ is the stretching exponent. Since $\varepsilon/\dot{\varepsilon}=t$ on each monotonic branch, the material-time argument and the scalar functional form used in the original fits are unchanged, while the sign of the viscous stress follows directly from $\dot{\varepsilon}$.

Finally, for polymer glasses such as PC and PMMA, one observes strain hardening at large deformation that arises from the finite extensibility and progressive alignment of polymer chains. We describe this contribution using a Langevin-based network model \cite{kroeger2015simple,nelson2026shear,hoy_2007strain_hardenig,arruda1993evolution,holopainen_2013modeling}, which captures the dominant entropic resistance to chain stretching. More detailed molecular mechanisms, including entanglement survival, chain orientation, and convective constraint release (CCR), can be incorporated through a strain-dependent evolution of load-bearing tube constraints within a nonaffine framework \cite{nichetti2026laos}. The resulting hardening stress is \cite{jordan2008mechanical,anand2006modeling},
\begin{equation}
\sigma_{\mathrm{hard}}(\varepsilon) =
E_R f(\lambda_{\varepsilon})
(\lambda_m/\lambda)
\mathcal{L}^{-1}
\left(\frac{\lambda}{\lambda_m}\right),
\end{equation}
where $E_R$ is the strain-hardening modulus of the polymer network,
$f(\lambda_{\varepsilon})=
\lambda_{\varepsilon}^{2}-\lambda_{\varepsilon}^{-1}$, and
\[
\lambda=
\sqrt{
\frac{\lambda_{\varepsilon}^{2}
+2\lambda_{\varepsilon}^{-1}}{3}
},
\qquad
\lambda_{\varepsilon}=\exp(\varepsilon),
\]
is the positive effective chain stretch used in the Eight-Chain model
\cite{jordan2008mechanical,anand2006modeling}. Although the axial
Hencky strain $\varepsilon$ is signed, $\lambda$ is positive by
construction, so that the inverse Langevin function depends on the
chain-stretch ratio $\lambda/\lambda_m$ rather than directly on the
signed axial strain. Here, $\mathcal{L}^{-1}(x)$ denotes the inverse
Langevin function \cite{kroeger2015simple}. An approximation is
$\mathcal{L}^{-1}(x)\approx 3x/(1-x^2)+x^3$, which captures the rapid
increase in stress as $\lambda$ approaches the finite-extensibility
limit $\lambda_m$.

The same axial stretch $\lambda_{\varepsilon}=\exp(\varepsilon)$ and
Hencky strain introduced above are used in the network contribution.
Thus, the elastic, viscous, and strain-hardening terms are embedded in
a common finite-deformation uniaxial kinematic description. The
Eight-Chain term retains its standard finite-extensibility dependence
through the effective chain stretch $\lambda$ and is interpreted, in
the present uniaxial specialization, as a contribution to the same
equivalent Cauchy stress used for the elastic and viscous terms.

The total signed equivalent Cauchy stress is then given by
\begin{equation}\label{final_stress}
\sigma_{\mathrm{tot}}(\varepsilon)=\sigma_{\mathrm{el}}(\varepsilon)+\sigma_{\mathrm{visc}}(\varepsilon)+\sigma_{\mathrm{hard}}(\varepsilon).
\end{equation}
All three terms in Eq.~\eqref{final_stress} are therefore contributions to the same signed Cauchy-stress measure. Their kinematic arguments are finite-deformation quantities associated with the same uniaxial loading: the elastic contribution uses the signed distortional Hencky measure $\Gamma$ together with its equivalent magnitude $\Gamma_{\mathrm{eq}}$, the viscous contribution uses the signed Hencky strain and rate obtained from $\mathbf{D}=\mathrm{sym}(\dot{\mathbf{F}}\mathbf{F}^{-1})$, and the network contribution uses the finite chain stretch constructed from $\lambda_{\varepsilon}=\exp(\varepsilon)$. For the monotonic proportional loading paths considered here, this finite-deformation reformulation retains the scalar functional forms used to obtain the stress--strain curves while placing the three contributions at the same kinematic and stress-description level.
Here total stress is expressed as the sum of elastic, viscous, and
strain-hardening contributions. This decomposition does not imply that the
underlying microscopic mechanisms are independent or that the transitions
between them are abrupt. Rather, it represents a coarse-grained constitutive
description in which the relative importance of the different mechanisms
evolves continuously with deformation. At small strains, the response is
dominated by nonaffine elastic deformation. With increasing strain,
irreversible structural rearrangements progressively contribute through
viscous relaxation. At large strains, network stretching becomes increasingly
important and gives rise to strain hardening. The additive form therefore
provides a first-order approximation for describing the different deformation
regimes within a tractable constitutive picture. In metallic glasses, the response is dominated by elastic loading, yielding, and flow, whereas in polymer glasses, the additional Langevin term leads to pronounced strain hardening at large deformation. Equation \eqref{final_stress} therefore represents a parallel additive decomposition of the different stress contributions. The first term decays to zero at large strain magnitude, the second term approaches a constant, and the strain-hardening contribution rises sharply as the chain stretch approaches its extensibility limit. In the case of metallic glasses, the stress-strain response is well described by the combined elastic and viscous contributions alone, as strain hardening is typically absent.

Only at this stage, when comparing with the compression datasets plotted with positive axes, do we drop the common minus sign from stress and strain. We therefore define the plotted quantities
\begin{equation}
\varepsilon_{\mathrm{plot}}=s\varepsilon,
\qquad
\sigma_{\mathrm{plot}}=s\sigma,
\end{equation}
with $s=-1$ for compression and $s=+1$ for tension. For the monotonic compression experiments considered in the figures, $\varepsilon_{\mathrm{plot}}$ and $\sigma_{\mathrm{plot}}$ are simply the positive strain and stress magnitudes conventionally reported.

The constitutive equation~\eqref{final_stress} recovers the appropriate limiting behavior in the different material classes. In the small-strain limit, the signed response is linear elastic, $\sigma\simeq E\varepsilon$, or equivalently $\sigma_{\mathrm{plot}}\simeq E\varepsilon_{\mathrm{plot}}$. For metallic glasses, for which $\sigma_{\mathrm{hard}}=0$, the elastic contribution vanishes at large deformation and the signed viscous response approaches $\sigma\simeq\eta\dot{\varepsilon}$; in the plotted convention this is $\sigma_{\mathrm{plot}}\simeq\eta\,s\dot{\varepsilon}$. For polymer glasses, the finite-extensibility contribution becomes dominant at large deformation and produces strain hardening.

The Poisson ratios used in the present analysis were taken from the
literature, with $\nu=0.36$ for metallic glass~\cite{lu2003deformation},
$\nu=0.38$ for PC~\cite{mulliken2006mechanics},
$\nu=0.35$ for PMMA~\cite{mulliken2006mechanics},
and $\nu=0.38$ for epoxy~\cite{jordan2008mechanical}.

\section{Results and discussion}

\begin{figure}[t]
\centering
\includegraphics[width=\linewidth]{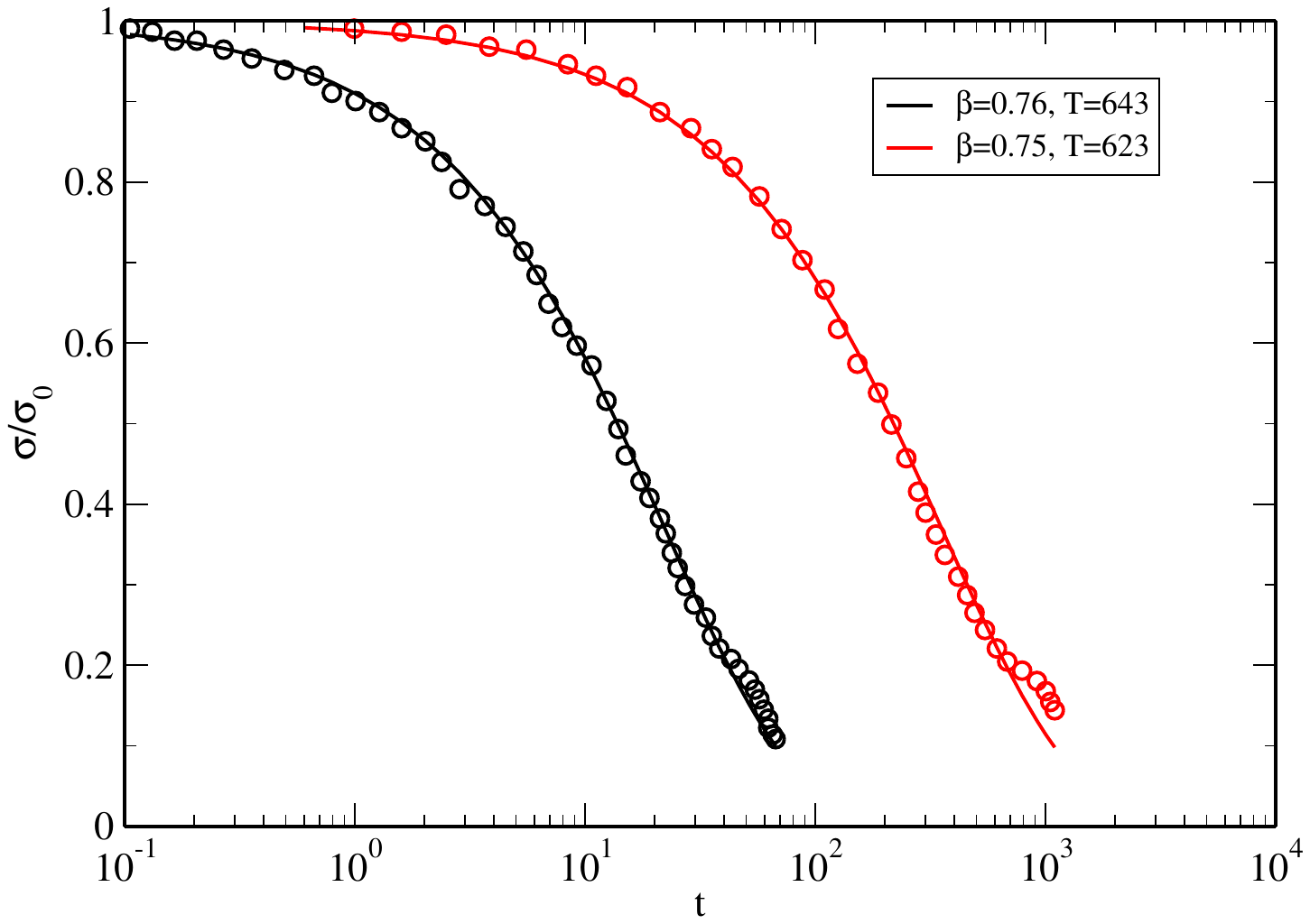}
\caption{Plot of the normalized stress as a function of time at two temperatures, $T = 643$ and $623~\mathrm{K}$. The symbols represent experimental data, while the solid lines correspond to stretched exponential fits with exponents $\beta = 0.76$ and $0.75$, respectively.}
\label{stress_time}
\end{figure}

\begin{figure}[t]
\centering
\includegraphics[width=\linewidth]{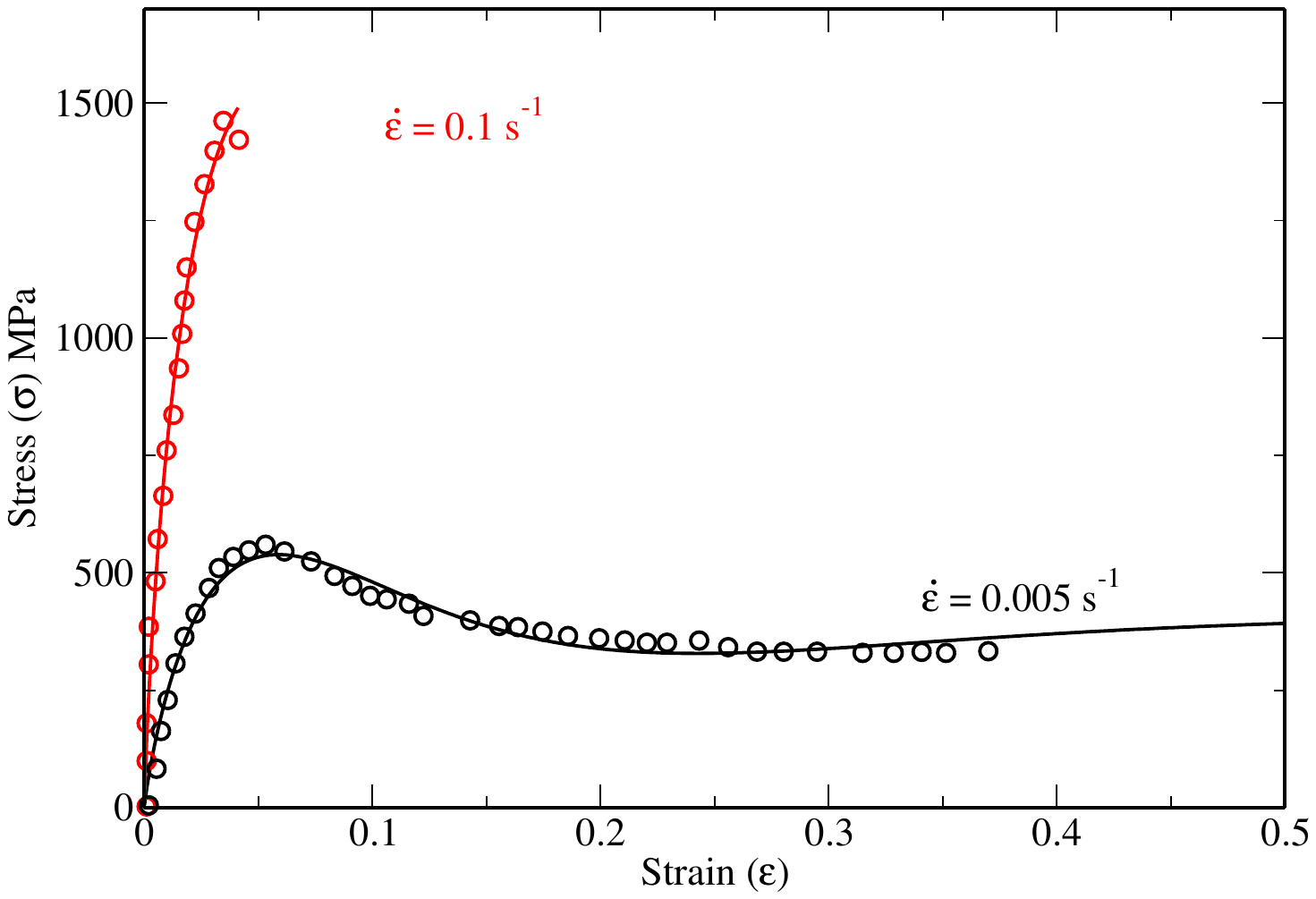}
\caption{Stress--strain curves of a metallic glass at two strain rates,
$\dot{\varepsilon}=0.1$ and $0.005~\mathrm{s^{-1}}$, at
$T=643~\mathrm{K}$. The solid lines are theoretical predictions and the
symbols are experimental data from Ref.~\cite{lu2003deformation}.}
\label{MG_stress}
\end{figure}

We now compare the theoretical results with experimental measurements on metallic and polymer glasses. When deformation is not dominated by shear banding, i.e., at moderate strain rates, the stress-strain behavior typically exhibits a pronounced overshoot, characterized by a maximum stress followed by a yielding regime. At larger strains, this response evolves toward viscous (Newtonian) flow, particularly in metallic glasses. In polymers, the yielding plateau can be observed at intermediate strains, when the deformations are small enough to preserve the original entanglement network. Once the deformations are large enough to strain or destroy the entanglements, the system enters the strain-hardening region. The magnitude and position of the stress overshoot depend on temperature and strain rate, making it a stringent test for theoretical descriptions of deformation.

The stretching exponent $\beta$ is first extracted independently from stress-relaxation measurements (Fig.~\ref{stress_time}) and subsequently used without further adjustment in the nonlinear stress-strain calculations. We observe from Fig.~\ref{stress_time} that, for the metallic glass, the $\beta$ value changes only slightly, from 0.76 to 0.75, at and above the glass transition temperature. Using $\beta$, along with experimentally known quantities, we calculate the stress-strain response at a given temperature $T=643$~K for strain rates $\dot{\varepsilon}=0.005$, $0.1~\mathrm{s}^{-1}$ with glass transition temperature $625 $~K \cite{peker1993highly,lu2003deformation}, determining the characteristic timescales $\tau_c$ and $\tau_v$. As shown in Fig.~\ref{MG_stress},  the theoretical predictions (solid lines) are compared with the experimental data (open circles) taken Ref.~\cite{lu2003deformation}, where the stress increases linearly at small strain, followed by an overshoot. The stress overshoot emerges from the competition between a nonaffine elastic instability associated with connectivity loss and the progressive buildup of viscous stress. Once the elastic instability is reached, the stress attains a maximum $\sigma_{\mathrm{max}}$ and subsequently decreases with increasing strain, while the viscous contribution grows monotonically toward the steady-state Newtonian plateau. Table~\ref{table} summarizes the model and experimental parameters. Most parameters are independently estimated from material
properties, experimental conditions, or relaxation data,
while $\tau_c$ provides the main adjustable timescale controlling
the onset of the elastic instability. 
The parameters $E$, $\eta$, $\tau_c$, $\tau_v$, $\beta$, and $E_R$ characterize the constitutive response and are listed in the table. The elastic modulus $E$, strain-hardening modulus $E_R$, and stretching exponent $\beta$ characterize the material response at the specified temperature and are treated as material parameters. The viscosity $\eta$ is an experimentally rate-dependent material property, while the relaxation times $\tau_c$ and $\tau_v$ are obtained from the constitutive fits and characterize the structural and viscous relaxation kinetics, respectively. The applied strain rate $\dot{\varepsilon}$ is an experimental control parameter. Although the dimensional cage-relaxation time $\tau_c$ varies with the applied strain rate, its variation is systematic rather than arbitrary. For a given material, the fitted values show that the dimensionless product \cite{lochhard1993rheological} $De_c=\dot{\varepsilon}\tau_c$, where $De_c$ denotes the cage-relaxation Deborah number that compares the $\tau_c$ with the imposed deformation timescale $\dot{\varepsilon}^{-1}$, remains approximately constant over the investigated strain-rate range. Specifically, $De_c \simeq 0.072$-$0.083$ for the metallic glass, $0.138$-$0.167$ for PC, $0.139$-$0.220$ for PMMA, and $0.144$-$0.218$ for epoxy. This behavior corresponds to $\tau_c\propto\dot{\varepsilon}^{-1}$ and suggests that the transferable quantity is the dimensionless relaxation parameter rather than the dimensional relaxation time itself.

\begin{figure}[h]
    \centering
    \includegraphics[width=0.98\linewidth]{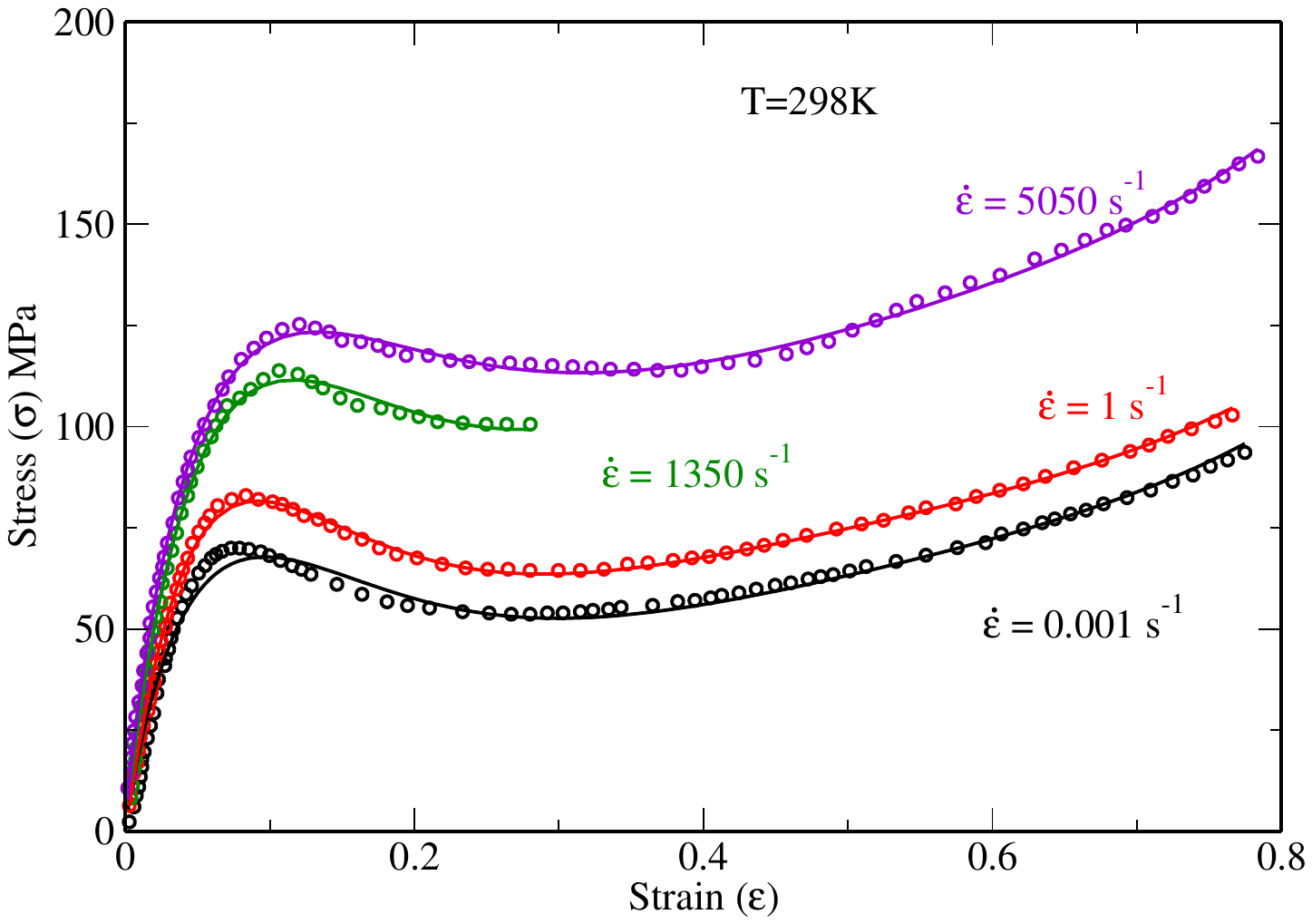}
    \caption{Stress-strain behavior of polycarbonate (PC) at temperature $T = 298~\mathrm{K}$ for different strain rates. The theoretical predictions are compared with experimental data obtained from Ref.~\cite{mulliken2006mechanics}.
}
    \label{PC_stress}
\end{figure}
\begin{figure}[t]
    \centering
    \includegraphics[width=0.98\linewidth]{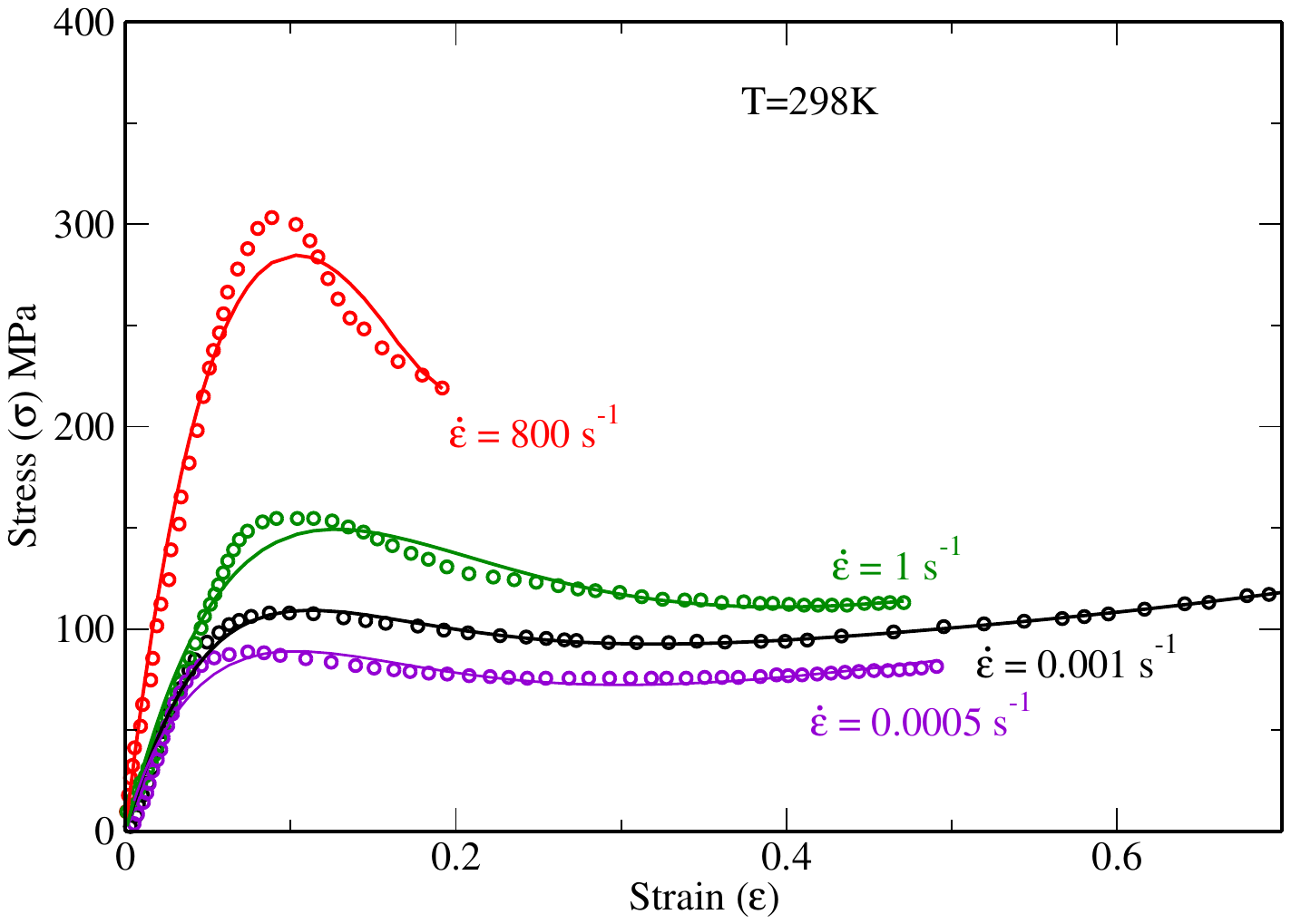}
    \caption{Stress-strain behavior of polymethyl methacrylate (PMMA) at temperature $T = 298~\mathrm{K}$ for different strain rates. The theoretical predictions are compared with experimental data obtained from Ref.~\cite{jordan2014mechanics,mulliken2006mechanics}.
}
    \label{PMMA_stress}
\end{figure}
\begin{figure}[t]
    \centering
    \includegraphics[width=0.98\linewidth]{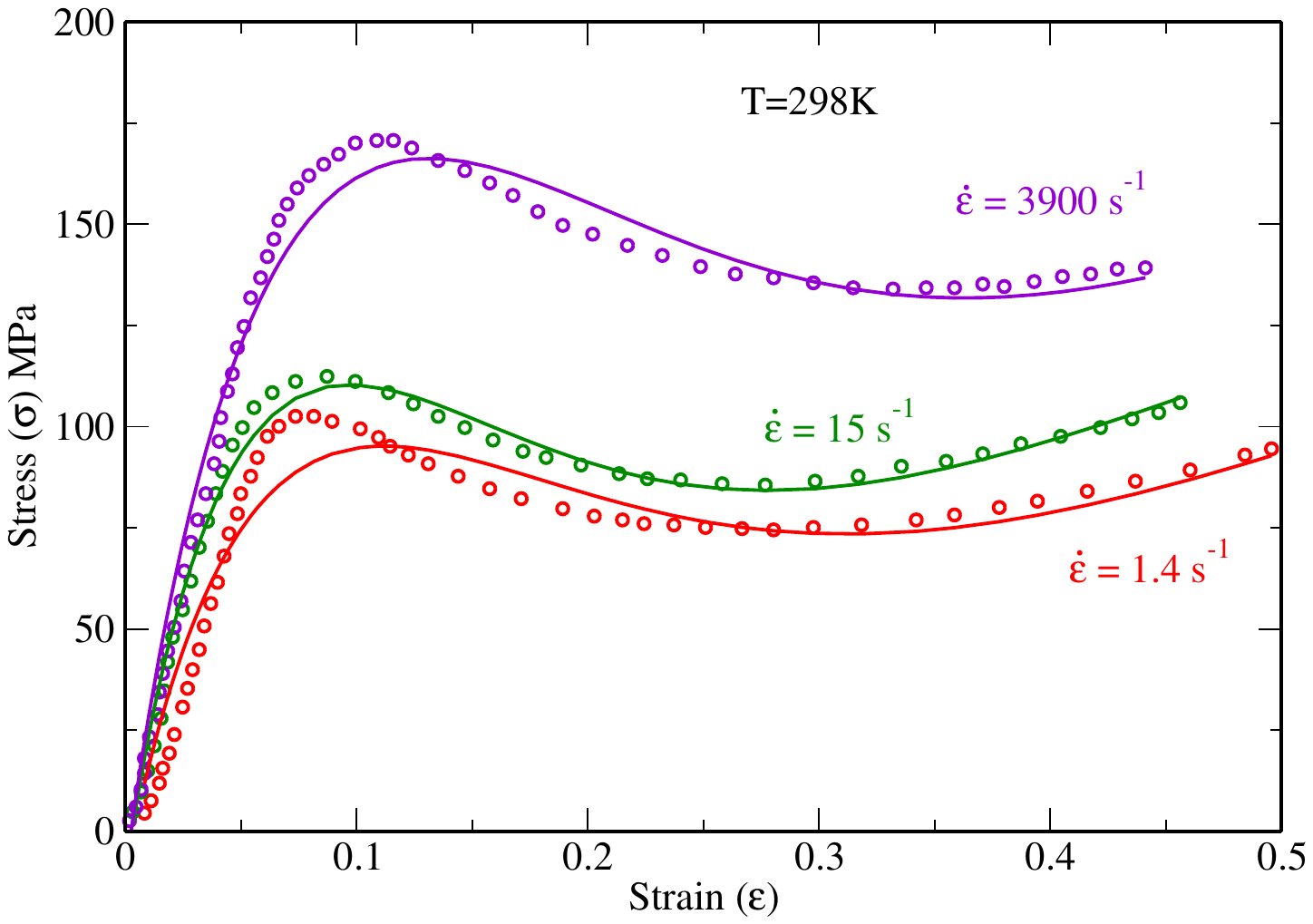}
    \caption{Stress-strain behavior of epoxy at temperature $T = 298~\mathrm{K}$ for three strain rates, $\dot{\varepsilon} = 1.4$, $15$ and $3900~\mathrm{s^{-1}}$ respectively. The theoretical predictions are compared with experimental data obtained from Ref.~\cite{jordan2008mechanical}.
}
    \label{epoxy_stress}
\end{figure}

\begin{figure}[]
    \centering
    \includegraphics[width=0.98\linewidth]{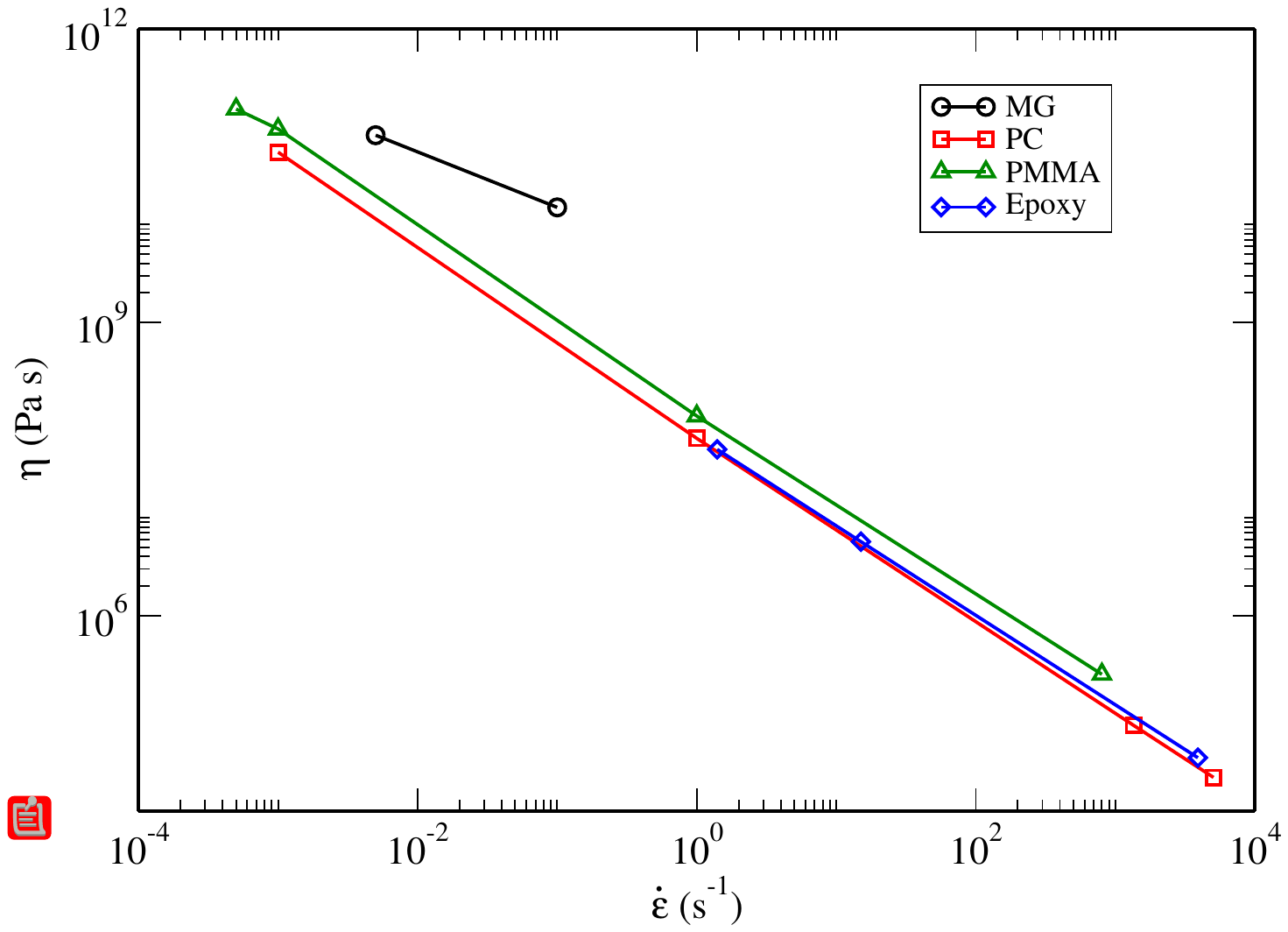}
    \caption{Log--log plot of the viscosity, \(\eta\) (Pa\,s), obtained from the linear viscous-flow region of the stress--strain curves as a function of strain rate, \(\dot{\varepsilon}\) (s\(^{-1}\)), for MG, PC, PMMA, and epoxy. Circles, squares, triangles, and diamonds represent the respective datasets. Linear trends indicate power-law scaling, with exponents \(-0.56\), \(-0.95\), \(-0.94\), and \(-0.91\) for MG, PC, PMMA, and epoxy, respectively.
}
    \label{viscosity}
\end{figure}

We next consider polymer glasses, where large-strain deformation is additionally influenced by chain stretching and finite extensibility \cite{kroeger2015simple}. At large strains, the hardening contribution is expressed in terms of the finite chain stretch $\lambda$ and the extensibility limit $\lambda_{\max}$, so that the large-deformation behavior of the Eight-Chain model is retained. The framework predicts the characteristic sequence of elastic loading, yielding, strain softening, plastic flow, and strain hardening observed experimentally.
Specifically, we consider polycarbonate (PC), polymethyl methacrylate (PMMA), and epoxy resins (Epon 826 epoxy resin cured with diethynolamine (DEA)). Resulting stress-strain curves at a given temperature and strain rate $\dot{\varepsilon}$ show the characteristic features observed experimentally and are compared with our theoretical predictions for glassy polymers. At small strain, the stress increases linearly with strain up to a maximum corresponding to the yield stress, defining the elastic regime, where the response is dominated by affine and weakly nonaffine elastic contributions. Beyond the yield point, the stress decreases with increasing strain, marking the strain softening regime, which is associated with the progressive loss of connectivity and the onset of irreversible microscopic rearrangements. At intermediate strain, the stress reaches a plateau corresponding to steady plastic flow, where the rate of structural rearrangements balances the applied deformation and the system flows at nearly constant stress. At larger strains, the stress increases again, signaling the strain hardening regime, which is generally attributed to increasing molecular orientation and stretching of polymer chains, as well as network constraints and the intrinsic strength and stiffness of covalent bonds that progressively resist further deformation.

\begin{table}[]
\caption{Fitting parameters used to describe the stress-strain response of metallic glass (MG) at $T = 643,\mathrm{K}$ and of polycarbonate (PC), poly(methyl methacrylate) (PMMA), and epoxy at $T = 298,\mathrm{K}$. Here, $E$ is the uniaxial elastic modulus, $\eta$ is the viscosity extracted from the linear viscous-flow region of the stress-strain curves, $\tau_c$ and $\tau_v$ are the structural and viscous relaxation times, $\beta$ is the stretching exponent, $E_R$ is the strain-hardening modulus with $\lambda_m = 1.82$, and $\dot{\varepsilon}$ is the applied strain rate, which is an experimental control variable.}
\vspace{0.5cm}
\centering
\resizebox{0.99\columnwidth}{!}{%
\begin{tabular}{lcccccccc}
\hline
Material & $E$ (GPa) & $\eta$ (Pa s) & $\tau_c$ (s) & $\tau_v$ (s) & $\beta$ & &$E_R$(MPa)& $\dot{\varepsilon}$ (s$^{-1}$) \\
\hline
MG & $8.1$ & $8.3\times 10^{10}$ & $16.5$ & $7.3$ & 0.76 & &$ $& $0.005$ \\
               & $12.5$ & $1.5\times 10^{10}$ & $0.72$ & $0.24$ & 0.76 & &$ $& $0.10$ \\
\hline
PC & 0.64 & $5.5\times 10^{10}$ & $138$ & $60$ & 0.85 & &$1.3$& $0.001$ \\
  & 0.67 & $6.5\times 10^{7}$ & $0.14$ & $0.05$ & 0.85 & &$1.4$& $1$ \\
  & 0.90 & $7.5 \times 10^{4}$ & $1.1 \times 10^{-4}$ & $6.1 \times 10^{-5}$ & 0.85 & &$1.8$& $1350$ \\
  & 0.71 & $2.2 \times 10^{4}$ & $3.3\times10^{-5}$ & $1.3 \times10^{-5}$ & 0.80 & &$1.7$& $5050$ \\
\hline
PMMA & 0.70 & $1.5\times 10^{11}$ & $278$ & $136$ & 0.85 & &$1.6$& $0.0005$ \\
  & 0.71 & $9.5\times 10^{10}$ & $164$ & $64$ & 0.85 & &$1.4$& $0.001$ \\
  & 1.13 & $1.1\times 10^{8}$ & $0.22$ & $0.10$ & 0.85 & &$1.9$& $1$ \\
  & 1.67 & $2.5\times10^{5}$ & $2.4\times10^{-4}$ & $3.8\times10^{-5}$ & 0.85 & &$ $& $800$ \\
\hline
Epoxy & 0.88 & $5.5\times 10^{7}$ & $0.11$ & $0.07$ & 0.85 & &$2.8$& $1.4$ \\
  & 1.21 & $5.7 \times 10^{6}$ & $9.6 \times 10^{-3}$ & $6.7 \times 10^{-3}$ & $0.85$ & &$3.9$& $15$ \\
  & 1.44 & $3.5\times10^{4}$ & $5.6\times10^{-5}$ & $4.3\times10^{-5}$ & $0.85$ & &$3.9$& $3900$ \\
\hline\end{tabular}}
\label{table}
\end{table}

In Figs.~\ref{PC_stress} and \ref{PMMA_stress} we plot the stress-strain curves for PC and PMMA at $T=298$~K for several strain rates, extending up to large deformation with glass transition temperature $423$~K and $388$~K respectively. Open symbols represent experimental data obtained from Ref.~\cite{jordan2014mechanics,mulliken2006mechanics}, while solid lines correspond to the theoretical predictions. The fitting parameters used to match the experimental data are summarized in Table~\ref{table}. In the case of PC, the yield stress increases and shifts to higher strain with increasing strain rate, reflecting the rate-dependent resistance to deformation. This is followed by a strain softening regime and a plastic flow plateau in the range $\varepsilon \approx 0.2$-$0.4$, depending on the applied strain rate. At larger strains, strain hardening becomes dominant, particularly at higher strain rates, indicating enhanced chain stretching and network constraints. A similar qualitative behavior is observed for PMMA, as shown in Fig.~\ref{PMMA_stress}; however, PMMA exhibits comparatively weaker strain hardening than PC, which can be attributed to its lower degree of chain entanglement and reduced molecular orientation capacity under deformation.


In Fig.~\ref{epoxy_stress} we plot the stress-strain response of epoxy at $T = 298~\mathrm{K}$ for different strain rates, $\dot{\varepsilon} = 1.4$, $15$, and $3900~\mathrm{s^{-1}}$. The theoretical predictions are compared with experimental data from Ref.~\cite{jordan2008mechanical}, using an estimated glass transition temperature $T_g \approx 350~\mathrm{K}$ for the epoxy system. In comparison to PC and PMMA, the yield point in epoxy appears sharper at lower strain rates, which can be attributed to its highly cross-linked network structure and reduced chain mobility. Despite this, the model captures the overall stress evolution in good agreement with experiments, reproducing the elastic, plastic flow, and strain hardening regimes reasonably well. The corresponding fitting parameters are summarized in Table~\ref{table}.

Figure~\ref{viscosity} shows the viscosity, $\eta$, extracted from the linear viscous-flow region of the stress--strain curves as a function of strain rate, $\dot{\varepsilon}$, for MG, PC, PMMA, and epoxy. For all materials, the viscosity decreases with increasing strain rate, indicating a non-Newtonian shear-thinning response. The metallic glass exhibits the highest viscosity over the investigated strain-rate range, whereas the polymeric systems show comparatively lower viscosities. On a log--log scale, the viscosity exhibits linear dependence on strain rate, with distinct slopes for different materials, suggesting power-law scaling behavior. The continuous increase in viscosity with decreasing strain rate indicates that the systems have not yet reached the low-rate Newtonian plateau; instead, a steady-state viscosity may emerge only at strain rates lower than those explored in the present study.

\section{Conclusions}
We have developed a microscopically informed constitutive framework for nonlinear deformation in metallic and polymer glasses based on nonaffine elasticity and irreversible structural relaxation. The framework identifies yielding as a nonaffine elastic instability driven by strain-induced connectivity loss and shows that stress overshoot is the direct manifestation of its competition with viscous relaxation. Using a single framework, the model quantitatively captures elastic loading, yielding, plastic flow, and strain hardening across a broad range of amorphous materials and deformation conditions. These results establish a direct connection between microscopic relaxation dynamics and macroscopic mechanical response, providing a  description of nonlinear deformation in glasses. For polymer glasses, strain hardening is described here at the level of finite chain extensibility. A more microscopic treatment that explicitly incorporates entanglement survival, chain orientation, and convective constraint release within a nonaffine framework has recently been proposed for nonlinear polymer rheology and LAOS \cite{nichetti2026laos}. Extending the present constitutive framework along these lines represents a promising direction for future work.

\section*{Acknowledgements}
AS and AZ are grateful for the funding from the European Union through Horizon Europe ERC Grant number: 101043968, “Multimech”. We acknowledge useful discussions with Dr. Dario Nichetti.

\bibliographystyle{elsarticle-num}
\bibliography{manuscript}

\end{document}